\documentclass[10pt,aps,prd,twocolumn,showpacs,nofootinbib,superscriptaddress]{revtex4-1}

\pdfoutput=1

\usepackage{graphicx}
\usepackage{subfigure}
\usepackage{amsmath}
\usepackage{amssymb}
\usepackage{bm} 
\usepackage{slashed}

\usepackage[usenames]{color}

\newcommand{\LLsum}{\sum_{s = \pm 1} \! \sum_{n=0}^{\nu \leq \nu_{\rm max}} \!}
\newcommand{\LLsumFT}{\sum_{s = \pm 1} \! \sum_{n=0}^\infty}
\newcommand{\beq}{\begin{equation}}
\newcommand{\eeq}{\end{equation}}
\newcommand{\bqa}{\begin{eqnarray}}
\newcommand{\eqa}{\end{eqnarray}}
\def\sumint{\hbox{\large$\sum$}\!\!\!\!\!\!\!\!\int}
\newcommand{\Tr}{\rm Tr}

\begin{document}

\title{Bulk Properties of a Fermi Gas in a Magnetic Field}

\author{M. Strickland}
\affiliation{Physics Department, Gettysburg College, Gettysburg, PA 17325 United States}

\author{V.~Dexheimer}
\affiliation{Physics Department, Gettysburg College, Gettysburg, PA 17325 United States}
\affiliation{Depto de F\'{\i}sica - CFM - Universidade Federal de Santa
Catarina  Florian\'opolis\\SC - CP. 476 - CEP 88.040 - 900 - Brazil}

\author{D. P. Menezes}
\affiliation{Depto de F\'{\i}sica - CFM - Universidade Federal de Santa
Catarina  Florian\'opolis\\SC - CP. 476 - CEP 88.040 - 900 - Brazil}


\date{\today}

\begin{abstract}
We calculate the number density, energy density, transverse pressure, longitudinal pressure,
and magnetization of an ensemble of spin one-half particles in the presence of a homogenous background 
magnetic field.  The magnetic field direction breaks spherical symmetry causing the pressure transverse
to the magnetic field direction to be different than the pressure parallel to it.  We present explicit formulae 
appropriate at zero and finite temperature for both charged and uncharged particles including the effect of the 
anomalous magnetic moment.  We demonstrate that the resulting expressions satisfy the canonical relations, 
$\Omega =  - P_\parallel$ and $P_\perp = P_\parallel - M B$, with $M = - \partial \Omega/\partial B$ being the 
magnetization of the system.  We numerically calculate the resulting pressure anisotropy for a gas of protons 
and a gas of neutrons and demonstrate that the inclusion of the anomalous magnetic increases the level of 
pressure anisotropy in both cases.
\end{abstract}

\maketitle

\section{Introduction}

The determination of the bulk properties of a Fermi gas in the presence of a magnetic field is 
important for understanding neutron stars and the early-time dynamics 
of the quark gluon plasma created in relativistic heavy ion collisions.  In the presence of a uniform magnetic 
field, both the matter and the field contributions to the spacelike components of the 
energy-momentum tensor become anisotropic.  The degree 
of pressure anisotropy increases as the magnitude of the magnetic field increases.  In this paper we revisit the 
calculation of the bulk properties of a Fermi gas of spin one-half particles in a uniform magnetic field with the goal 
of unambiguously determining the pressure anisotropy from first principles including the effect of 
the anomalous magnetic moment.  

As mentioned above, there is currently considerable interest in the behavior of matter in the presence of
high magnetic fields.  Neutron stars, for example, are known to possess high magnetic fields.  More 
specifically, magnetars \cite{Duncan:1992hi,1992AcA....42..145P,%
1993ApJS...88..529T,1996ApJ...473..322T,1999ApJ...519L..77M,Ibrahim:2003ev,2006puas.book.....L}
are believed to have surface magnetic fields as strong as $10^{14}-10^{15}$ Gauss.
Based on such surface magnetic fields, one could expect magnetic fields in the interior of magnetars to be on
the order of $10^{16}-10^{19}$ Gauss.   There have been many previous studies of the 
effect of magnetic fields on neutron stars and magnetars focusing on the effect of magnetic fields on
the equation of state of the matter composing the star including hadronic matter, quark matter, and hybrid
stars composed of hadronic matter with a quark matter core \cite{Canuto:1969ct,Canuto:1969cs,%
Canuto:1969cn,Chiu:1968zz,1977FCPh....2..203C,%
BlandfordHernquist1982,1989ApJ...342..958F,1991ApJ...374..652A,1991ApJ...383..745L,%
1992AnPhy.216...29F,1993ApJ...416..276R,Chakrabarty:1996te,Chakrabarty:1997ef,%
Bandyopadhyay:1997kh,1998ApJ...502..847T,Chaichian:1999gd,YuanZhang1999,Broderick:2000pe,%
Broderick:2001qw,Mao:2001fq,Suh:2001tr,Bednarek:2002hb,Martinez:2003dz,PerezMartinez:2005av,%
PerezMartinez:2007kw,Felipe:2007vb,Menezes:2008qt,Rabhi:2008je,Huang:2009ue,Menezes:2009uc,%
Rabhi:2009ih,Ferrer:2010wz,Paulucci:2010uj,Orsaria:2010xx,Sinha:2010fm,Dexheimer:2011pz,%
Ryu:2011gq,2011PhRvC..83f5805A,Lopes:2012nf,Dexheimer:2012qk,Avancini:2012ee}.  

Among these references some authors have simply assumed that the system continues to be describable in 
terms of an energy density and an isotropic pressure derivable from standard thermodynamic relations, while other
authors have included the fact that the background magnetic field 
breaks the spherical symmetry of the system.  The breaking of the spherical symmetry has two distinct
contributions:  (i) the matter contribution to the energy-momentum tensor and (ii) the field contribution to 
the energy-momentum tensor.  For charged particles the presence of a magnetic field 
causes the pressure transverse to and longitudinal to the local magnetic field direction to be different, with 
the level of pressure anisotropy increasing monotonically with the magnitude of the magnetic field.  The
same occurs for uncharged particles that have a non-vanishing anomalous magnetic moment as we will 
demonstrate.

There have been dynamical models of neutron stars which have attempted to include the effect of 
high magnetic fields on the three-dimensional structure of neutron stars \cite{1954ApJ...119..407F,%
Bocquet:1995je,Rezzolla:1999he,Cardall:2000bs,Kiuchi:2007pa,Lander:2012iz}.  Some of these studies have 
self-consistently included modifications of the general relativistic metric necessary to describe the breaking of 
spherical symmetry by the neutron star's magnetic field.  However, to the best of our knowledge there has not 
been a study which has simultaneously included the general relativity aspects, effects of magnetic fields on
the equation of state, and effects of pressure anisotropy on the static and dynamical properties of a
high-magnetic-field neutron star.  In order to complete this program it is necessary to first understand
all sources of pressure anisotropy due to magnetic fields.  

Another area in which there has been a significant amount of attention focused on the behavior of
matter subject to high magnetic fields is the consideration of the first fm/c after the collision of two
high-Z ions in a relativistic heavy ion collision.  Because of the large number of protons in the colliding
nuclei, magnetic fields on the order of $10^{18} - 10^{19}$ Gauss are expected to be generated at early 
times after the initial nuclear impact \cite{Fukushima:2008xe,Skokov:2009qp,Fukushima:2010vw,%
Voronyuk:2011jd,Deng:2012pc}.  The existence of such high magnetic  fields prompted many research
groups to study how the finite temperature deconfinement and chiral phase transitions are affected 
by the presence of a background magnetic field.  These studies have included direct numerical investigations 
using lattice quantum chromodynamics (QCD) \cite{D'Elia:2010nq,D'Elia:2011zu,Bali:2011qj,Bali:2012zg} 
and theoretical investigations using a variety of methods including, for example, perturbative QCD studies,
model studies, and string-theory inspired anti-de Sitter/conformal field theory (AdS/CFT) correspondence 
studies \cite{Agasian:2008tb,Fraga:2008qn,Mizher:2010zb,Fukushima:2010fe,Gatto:2010qs,Gatto:2010pt,%
Preis:2010cq,Preis:2011sp,Andersen:2011ip,Erdmenger:2011bw,Gorbar:2011jd,Skokov:2011ib,Kashiwa:2011js,%
Fraga:2012fs,Fraga:2012ev,Fraga:2012rr,Andersen:2012dz,Shovkovy:2012zn,Preis:2012fh,%
Ferrari:2012yw,Fayazbakhsh:2012vr,Fukushima:2012xw,dePaoli:2012,Alexandre:2000jc}.  

In order to have more a comprehensive understanding of the behavior of matter in a background 
magnetic field, we begin with the basics and study Fermi gases consisting of charged and uncharged spin 
one-half particles including the effect of the anomalous magnetic moment.   
Many of the results obtained here are already available in the literature; however, the results for the
transverse pressure including the effect of the anomalous magnetic moment have not appeared previously.  
For sake of completeness, we present the results for all of the components of the matter contribution 
to the energy-momentum tensor with and without anomalous magnetic moment as a point of reference
for future applications.  In this paper we consider systems at both zero and finite temperature.
For zero temperature systems, we demonstrate by explicit 
calculation that the grand potential $\Omega = \epsilon - \mu n = - P_\parallel$ where $\epsilon$ is the energy 
density, $n$ is the number density, $P_\parallel$ is the pressure along the direction of the background
magnetic field, and $\mu$ is the chemical potential.  For finite temperature systems one also finds that $\Omega 
= - P_\parallel$.

We then show that, both with and without anomalous magnetic moment, the resulting expressions satisfy the canonical relation $P_\perp = P_\parallel - M B$, where $P_\perp$ is the pressure 
transverse to the magnetic field direction and $M = - \partial \Omega/\partial B$ is the magnetization of the 
system.   Evaluating the resulting expressions numerically, we demonstrate that the magnitude 
of the pressure anisotropy is larger when one takes into account the anomalous magnetic moment, however, 
as the temperature of the system increases the pressure anisotropy decreases. 

The structure of the paper is as follows.  In Sec.~\ref{sec:general}
we  introduce the basic formulae necessary to calculate the bulk properties of an ensemble of particles 
using quantum field theory.  In
Sec.~\ref{sec:charged} we present the resulting formulae for charged particles with and without 
anomalous magnetic moment.  In Sec.~\ref{sec:uncharged} we present the corresponding formulae 
for uncharged particles.   In Sec.~\ref{sec:results}
we compare the numerical evaluation of the transverse and longitudinal pressures.
In Sec.~\ref{sec:conclusions} we present our conclusions and an outlook for the future.
Finally, in Apps.~\ref{app:tmunubasics} and \ref{app:tmunuderivation} we present a quantum field theory 
derivation of the necessary components of the energy-momentum tensor for charged and uncharged particles.

\section{Generalities}
\label{sec:general}

In the presence of fields, the energy-momentum tensor can be decomposed into matter 
and field contributions
\begin{equation}
T^{\mu\nu} = T^{\mu\nu}_{\rm matter} + T^{\mu\nu}_{\rm fields} \, .
\end{equation}

If there is only a background magnetic field $B$ pointing along the $z$-direction, then the 
field contribution to the energy-momentum tensor takes the form $T^{\mu\nu}_{\rm fields}
= {\rm diag}(B^2/2,B^2/2,B^2/2,-B^2/2)$.\footnote{This is the form in Heaviside-Lorentz
natural units.  In Gaussian natural units, when converting the magnetic field to GeV$^2$,
the magnetic field is increased by a factor of $\sqrt{4\pi}$ 
and the components of the energy momentum-tensor   
are divided by a factor of $4 \pi$ to 
compensate, e.g. $\epsilon_B = B^2/8\pi$.}  Since this contribution is well-understood,
we  
do not spend more time discussing it in this paper.  Instead, we  
focus on $T^{\mu\nu}_{\rm matter}$ for a system  
composed of spin one-half fermions.  In what follows, the bulk 
properties of the system (energy density, transverse pressure, etc.) are understood to specify 
the components of  $T^{\mu\nu}_{\rm matter}$ in the local rest frame of the system.

The matter contribution to the bulk properties of a system can be expressed in terms of the one-particle
distribution function $f$.  We  m
consider a single particle type with mass $m$ and charge $q$ and
sum over the spin polarizations.  The results obtained can be straightforwardly extended to a system 
consisting of multiple particle types.  
We present a derivation of the necessary components of
the energy-momentum tensor in Apps.~\ref{app:tmunubasics} and \ref{app:tmunuderivation}.  
Summarizing the results, one finds that the local rest frame number density, energy density, longitudinal 
pressure, and transverse pressure can be expressed in terms of the following integrals of the one-particle
distribution function
\begin{eqnarray}
n &=& \sum_s \int_k f \, ,
\label{ndens}
\\
\epsilon &=& T^{00} = \sum_s \int_k E f \, ,
\label{edens}
\\
P_\parallel &=& T^{zz} = \sum_s \int_k \frac{k_z^2}{E} f \, ,
\label{ppar}
\\
P_\perp &=& \frac{1}{2}\left( T^{xx} + T^{yy} \right) 
\nonumber
\\
&=& \sum_s \int_k 
\frac{1}{E} 
\Biggl[ 
\frac{1}{2} \frac{k_\perp^2 \bar{m}(\nu)}{\sqrt{m^2 + k_\perp^2}}
-   s \kappa B \bar{m}(\nu)
 \Biggr] f \, ,
\label{pperp}
\end{eqnarray}
where we have singled out the $z$ (parallel) direction for future application, 
$\bar{m}^2(\nu) \equiv (\sqrt{m^2 + k_\perp^2} - s \kappa B)^2$,  
$k_\perp^2$ is the (discretized) transverse momentum,  $\sum_s$ represents a sum over spin polarizations,
$\kappa$ represents the anomalous magnetic moment,
and $\int_k$ is a properly normalized (sum-)integration over momenta which we will define
separately for charged and uncharged particles. 
For charged particles with vanishing anomalous magnetic moment, 
the expressions above were first derived in Ref.~\cite{Canuto:1969ct}.  For charged particles with 
finite anomalous magnetic moment the expressions for the number density and 
energy density above were first derived in Ref.~\cite{Chiu:1968zz}.  Here we extend
the treatment to include uncharged particles and independently compute the
transverse and longitudinal pressures in the case of finite anomalous magnetic moment.

We note that in order to include interactions, one should use the interaction-corrected expression for the particle's
dispersion relation.  In the mean-field approximation, this amounts to including corrections to the 
bare mass of the particle being considered, e.g. $m \rightarrow m^*$.   The resulting effective mass 
can depend on the chemical potential and temperature.  In what follows
we 
indicate the effective mass of the particle as $m$ assuming that interaction corrections could be absorbed into the mass.\footnote{In 
the following, spherical symmetry is broken by a uniform magnetic field.  Due 
to this, the effective mass could, in principle, also depend on the angle of particle momentum relative to the magnetic field
direction.  We do not take this possibility into account in this work.}

\section{Charged Particles}
\label{sec:charged}

In the presence of a uniform external magnetic field pointing in the $z$-direction, the transverse momenta of
particles with an electric charge $q$ are restricted to discrete Landau levels with $k_\perp^2 = 2 \nu |q| B$ where 
$\nu \geq 0$ is an integer \cite{LandauLifshitzQM} and one has 
\begin{equation}
\int_k \rightarrow \frac{|q| B}{(2\pi)^2} 
\sum_n \int_{-\infty}^\infty d k_z \, ,
\label{lsumform}
\end{equation}
where the sum over $n$ represents a sum over the 
discretized orbital angular momentum of the particle in the transverse plane.  For
spin one-half particles the orbital angular momentum $n$ is related to $\nu$ via
\cite{LandauLifshitzQM}
\begin{equation}
\nu = n + \frac{1}{2} - \frac{s}{2} \frac{q}{|q|} \, ,
\label{nudef}
\end{equation}
where $s=\pm1$ is the spin projection of the particle along the
direction of the magnetic field and $q$ is the charge.\footnote{The present calculation is valid only for spin one-half
particles. Spin zero, one and three-half particles, described respectively
by the Klein-Gordon, Proca and Rarita-Schwinger equations are affected differently by the magnetic field 
\cite{dePaoli:2012, dePaoli2:2012}.}

An additional consequence of the quantization is that the total energy of a charged particle becomes
quantized \cite{ternov1966}
\begin{eqnarray}
E &=& \sqrt{k_z^2 + ((m^2 + 2 \nu |q| B)^{1/2} - s \kappa B)^2} \, , \nonumber \\
&=& \sqrt{k_z^2 + \bar{m}^2(\nu)} \, ,
\label{benergy}
\end{eqnarray}
where $\kappa = \kappa_i \mu_N$ with $\kappa_i$ being the coupling strength 
for the anomalous magnetic moment times the magneton, and 
$\bar{m}^2(\nu) \equiv (\sqrt{m^2 + 2 \nu |q| B} - s \kappa B)^2$.

\subsection{Zero temperature}

At zero temperature the one-particle distribution function is given by a Heaviside theta
function
\begin{equation}
f(E) = \Theta(\mu - E) \, ,
\end{equation}
where $\mu$ is the chemical potential (Fermi energy).  

\subsubsection{Zero anomalous magnetic moment}

We begin by considering the case with no anomalous magnetic moment, i.e. $\kappa = 0$.
In terms of the chemical potential, $\mu$, the maximum $k_z$ is defined via (\ref{benergy})
\begin{equation}
k_{z,F}(\nu) = \sqrt{\mu^2 - 2 \nu |q| B - m^2} \, .
\label{kzF}
\end{equation}
In addition, in the sum over the Landau levels one must guarantee that the quantity under the
square root in (\ref{kzF}) is positive.  This requires $\bar{m}^2 \leq \mu^2$ which results in
\begin{equation}
\nu \le \nu_{\rm max} = \left\lfloor \frac{\mu^2 - m^2}{2|q|B} \right\rfloor ,
\label{numax} 
\end{equation}
where $\lfloor x \rfloor = {\rm max}\{ n \in \mathbb{Z} \; | \; n \leq x \}$ is the largest integer less
than or equal to $x$.

Using the above, we can write down an expression for the number density using (\ref{ndens}) and
(\ref{lsumform}) to obtain \cite{Chakrabarty:1996te,Broderick:2000pe}
\begin{eqnarray}
n &=& \frac{|q|B}{(2\pi)^2} \LLsum \int_{-\infty}^{\infty} dk_z \, \Theta(\mu - E) \, ,
\nonumber \\
&=& \frac{|q|B}{2\pi^2} \LLsum \int_{0}^{k_{z,F}} dk_z  \, ,
\nonumber \\
&=& \frac{|q|B}{2\pi^2} \LLsum \, k_{z,F}(\nu) \, .
\label{ndense}
\end{eqnarray}

Note that the upper limit on the $n$ sum is set in terms of the maximum Landau level
and that $\nu$ depends on $n$ and $s$ via Eq.~(\ref{nudef}).  Note that the $\kappa=0$ degeneracy
factor for a given Landau level is automatically taken into account by the dual sum over
spin and angular momentum.

Similarly, one can evaluate the energy density to obtain \cite{Chakrabarty:1996te,Broderick:2000pe}
\begin{eqnarray}
\epsilon &=& \frac{|q|B}{2\pi^2} \LLsum \int_{0}^{k_{z,F}} dk_z \, \sqrt{k_z^2 + \bar{m}^2(\nu)} \, ,
\nonumber \\
&=& \frac{|q|B}{4\pi^2} \LLsum \Biggl[ \mu \, k_{z,F}(\nu) 
\nonumber \\
&& \hspace{1.5cm} + \bar{m}^2(\nu) \log\left(\frac{\mu + k_{z,F}(\nu)}{\bar{m}(\nu)}\right) \Biggr] .
\label{edense}
\end{eqnarray}

Next, we consider the parallel pressure $P_\parallel$ and obtain \cite{Chakrabarty:1996te}
\begin{eqnarray}
P_\parallel &=& \frac{|q|B}{2\pi^2} \LLsum \int_{0}^{k_{z,F}} dk_z \, \frac{k_z^2}{\sqrt{k_z^2 + \bar{m}^2(\nu)}} \, ,
\nonumber \\
&=& \frac{|q|B}{4\pi^2} \LLsum \Biggl[ \mu \, k_{z,F}(\nu) 
\nonumber \\
&& \hspace{1.5cm} - \bar{m}^2(\nu) \log\left(\frac{\mu + k_{z,F}(\nu)}{\bar{m}(\nu)}\right) \Biggr] .
\label{ppare}
\end{eqnarray}
Note that using (\ref{ndense}), (\ref{edense}), and (\ref{ppare}) it is straightforward to see  
that $\epsilon + P_\parallel = \mu n$ and hence $\Omega = \epsilon - \mu n = - P_\parallel$.

Finally, we consider the transverse pressure $P_\perp$ and obtain
\begin{eqnarray}
P_\perp &=& \frac{|q|B}{4\pi^2} \LLsum \, 2 \nu |q| B \, \int_{0}^{k_{z,F}} dk_z \, \frac{1}{\sqrt{k_z^2 + \bar{m}^2(\nu)}} \, ,
\nonumber \\
&=& \frac{|q|^2B^2}{2\pi^2} \LLsum \, \nu \, \log\left(\frac{\mu + k_{z,F}(\nu)}{\bar{m}(\nu)}\right) .
\label{pperpe}
\end{eqnarray}

Numerically the results for $P_\parallel$ and $P_\perp$ are different for any value of $B$; however, they only become 
significantly different for very large $B$.  Using Eq.~(\ref{numax}), for example, we see that when $B > (\mu^2 - m^2)/2|q|$,
only the lowest Landau level contributes to the sums and one obtains 
\begin{equation}
\lim_{B \rightarrow \infty} 
P_\parallel = \frac{|q|B}{4\pi^2}\Biggl[ \mu \, k_F
- m^2 \log\left(\frac{\mu + k_F}{m}\right) \Biggr] ,
\label{pparib}
\end{equation}
where $k_F \equiv \sqrt{\mu^2 - m^2}$.  The transverse pressure on the other hand vanishes in this limit
\begin{equation}
\lim_{B \rightarrow \infty} P_\perp = 0 \, .
\end{equation}

A relationship between $P_\parallel$ and $P_\perp$ 
can be established by evaluating the magnetization of the system $M \equiv - \partial \Omega/\partial B = 
\partial P_\parallel/\partial B$ \cite{LandauLifshitzPitaevskii}.  Performing the necessary 
derivatives\footnote{Formally one should use left 
or right derivatives in the vicinity of magnetic field magnitudes where $\nu_{\rm max}$ changes under infinitesimal 
variation.}
of the parallel pressure one finds  $M = (P_\parallel - P_\perp)/B$.
Rearranging gives $P_\perp = P_\parallel - M B$ which is the canonical relationship one finds in the literature
between the transverse and longitudinal pressures.  

\subsubsection{Nonzero anomalous magnetic moment}

We now turn to the case of nonzero anomalous magnetic moment.  In this case the expressions
for $k_{z,F}$ and $\nu_{\rm max}$ must be adjusted to
\begin{equation}
k_{z,F} = \sqrt{\mu^2 - ((m^2 + 2 \nu |q| B)^{1/2} - s \kappa B)^2} \, ,
\label{akzf}
\end{equation}
\begin{equation}
\nu_{\rm max} = \left\lfloor \frac{(\mu + s \kappa B)^2 - m^2}{2 |q| B} \right\rfloor .
\label{anumax}
\end{equation}
With these two modifications Eqs.~(\ref{ndense}), (\ref{edense}), and (\ref{ppare})
are unchanged, but one should note that $\nu_{\rm max}$ now depends on the spin alignment $s$.

The transverse pressure, however, is modified when there is a non-vanishing anomalous
magnetic moment
\bqa
P_\perp &=&  
 \frac{|q| B^2}{2 \pi^2} \LLsum
 \Biggl[ 
\frac{|q| \nu \bar{m}(\nu)}{\sqrt{m^2 + 2 \nu |q| B}} -   s \kappa \bar{m}(\nu) \Biggr]
\nonumber \\
&& \hspace{2cm} \times
\int_{0}^{k_{z,F}} \! dk_z \, \frac{1}{\sqrt{k_z^2 + \bar{m}^2(\nu)}} \, .
\hspace{8mm}
\nonumber \\
&=&  
 \frac{|q| B^2}{2 \pi^2} \LLsum
 \Biggl[ 
\frac{|q| \nu \bar{m}(\nu)}{\sqrt{m^2 + 2 \nu |q| B}} -   s \kappa \bar{m}(\nu) \Biggr]
\nonumber \\
&& \hspace{3cm} \times
\log\left(\frac{\mu + k_{z,F}(\nu)}{\bar{m}(\nu)}\right)  \, .
\hspace{8mm}
\eqa
Evaluating the magnetization one obtains in this case 
\cite{Broderick:2000pe}\footnote{We note that there appear to be some typos in the expression contained
in Ref.~\cite{Broderick:2000pe}.}
\begin{eqnarray}
M &=& \frac{\partial P_\parallel}{\partial B} = \frac{P_\parallel}{B}
+ \frac{|q|B}{2\pi^2} \LLsum 
\nonumber \\
&& \hspace{-7mm} \times 
\Biggl[s \kappa  \bar{m}(\nu) - \frac{|q| \nu \bar{m}(\nu)}{\sqrt{m^2 + 2 \nu |q| B}} \Biggr] \!
\log\left(\frac{\mu + k_{z,F}(\nu)}{\bar{m}(\nu)}\right) \! . 
\hspace{5mm}
\nonumber \\
&=& \frac{P_\parallel}{B} - \frac{P_\perp}{B} \, .
\label{cpmag}
\end{eqnarray}
So one finds once again $P_\perp = P_\parallel - M B$.

\subsection{Finite temperature}

We now turn our attention to the case of a finite temperature ensemble of charged particles.  
In this case the distribution function is
\begin{equation}
f_\pm(E,T,\mu) = \frac{1}{e^{\beta(E \mp \mu)} + 1} \, ,
\end{equation}
where $f_+$ describes particles, $f_-$ describes anti-particles,  
and $\mu$ is the chemical potential.

\subsubsection{Zero anomalous magnetic moment}

We begin with the number density
\begin{equation}
n_\pm = \frac{|q|B}{(2\pi)^2} \LLsumFT  \int_{-\infty}^{\infty} dk_z \, f_\pm(E,T,\mu) \, ,
\end{equation}
recalling that $E =  \sqrt{k_z^2 + \bar{m}^2(\nu)}$ with $\bar{m}^2(\nu) = m^2 + 2 \nu |q| B$.
Introducing the variable $x = E\mp\mu$ we can rewrite $k_z = \sqrt{(x\pm\mu)^2 - 
\bar{m}^2(\nu)}$ and using $d k_z = (x\pm\mu) ((x\pm\mu)^2 - \bar{m}^2(\nu))^{-1/2} dx$ one
obtains
\begin{equation}
n_\pm = \frac{|q|B}{2\pi^2} \LLsumFT \int_{\bar{m}(\nu)\mp\mu}^{\infty} dx \, 
\frac{(x\pm\mu)f_\pm(x,T,0)}{\sqrt{(x\pm\mu)^2 - \bar{m}^2(\nu)}} \, .
\label{nft}
\end{equation}
Next, we consider the energy density.  Using the same change of variables as before, one obtains
\begin{equation}
\epsilon_\pm = \frac{|q|B}{2\pi^2} \LLsumFT \int_{\bar{m}(\nu)\mp\mu}^{\infty} dx \, 
\frac{(x\pm\mu)^2 f_\pm(x,T,0)}{\sqrt{(x\pm\mu)^2 - \bar{m}^2(\nu)}} \, .
\label{eft}
\end{equation}
Similarly, one obtains for the longitudinal pressure
\begin{eqnarray}
P_{\parallel,\pm} &=& \frac{|q|B}{2\pi^2} \LLsumFT 
\nonumber \\
&& \times \int_{\bar{m}(\nu)\mp\mu}^{\infty} dx \, 
\sqrt{(x\pm\mu)^2 - \bar{m}^2(\nu)} f_\pm(x,T,0) \, .
\nonumber \\
\label{pparft}
\end{eqnarray}
Finally, one obtains for the transverse pressure
\begin{equation}
P_{\perp,\pm} = \frac{|q|^2B^2}{2\pi^2} \! \LLsumFT \, \nu \!
\int_{\bar{m}(\nu)\mp\mu}^{\infty} \!\! dx \, 
\frac{f_\pm(x,T,0)}{\sqrt{(x\pm\mu)^2 - \bar{m}^2(\nu)}} \, .
\label{pperpft}
\end{equation}

Next we consider the magnetization obtained from $M = \partial P_{\parallel}/ \partial B$.
In order to do this we apply the fundamental theorem of calculus
\begin{eqnarray}
&& \frac{d}{dy} \int_{a(y)}^b dx \, g(x,y,\cdots) 
=
\nonumber \\
&& \hspace{4mm}
- a'(y) \, g(a(y),y,\cdots)  + \int_{a(y)}^b dx \, \frac{dg(x,y,\cdots)}{dy} \, .
\label{ftc}
\end{eqnarray}
Using this we can evaluate the derivative of the integral appearing on the second line 
of (\ref{pparft})
\begin{eqnarray}
&&\frac{\partial}{\partial B}\left( \int_{m(\nu)\mp\mu}^{\infty} dx \, 
\sqrt{(x\pm\mu)^2 - \bar{m}^2(\nu)} f_\pm(x,T,0) \right) =
\nonumber \\
&& \hspace{4mm}
- \bar{m}(\nu) \frac{\partial \bar{m}(\nu)}{\partial B} 
\int_{\bar{m}(\nu)\mp\mu}^{\infty} dx \, \frac{f_\pm(x,T,0)}{\sqrt{(x\pm\mu)^2 - \bar{m}^2(\nu)}} \, ,
\end{eqnarray}
where we have used the fact that in the case at hand the first term on the right-hand side of (\ref{ftc}) is zero.
Using $\bar{m} \, \partial \bar{m} / \partial B = \frac{1}{2} \partial \bar{m}^2/\partial B = |q| \nu$ we 
can obtain finally
\begin{equation}
M_\pm = \frac{\partial P_{\parallel,\pm}}{\partial B} = \frac{P_{\parallel,\pm}}{B} - \frac{P_{\perp,\pm}}{B} \, ,
\end{equation}
which is the canonical relation between the transverse pressure, the longitudinal pressure, and the magnetization.  
Rearranging we obtain $P_{\perp,\pm} = P_{\parallel,\pm} - M_\pm B$ between the perpendicular and parallel 
pressures at finite temperature in the case that there is no anomalous magnetic moment.

\subsubsection{Nonzero anomalous magnetic moment}

As was the case at zero temperature, when including the anomalous magnetic moment, the primary
thing that changes is the mass $\bar{m}^2(\nu) = (\sqrt{m^2 + 2 \nu |q| B} - s \kappa B)^2$.
With this change, the expressions for $n_{\pm}$, $\epsilon_{\pm}$, and $P_{\parallel,\pm}$ 
given in Eqs.~(\ref{nft}), (\ref{eft}), and (\ref{pparft}), respectively, are unchanged.  
For the transverse pressure, however, one must include additional terms
\begin{eqnarray}
P_{\perp,\pm} &=& \frac{|q| B^2}{2\pi^2} \! \LLsumFT \, 
\bar{m}(\nu) \left[\frac{|q| \nu}{\sqrt{m^2 + 2 \nu |q| B}} - s \kappa\right]
\nonumber \\
&& \hspace{1cm} \times
\int_{\bar{m}(\nu)\mp\mu}^{\infty} \!\! dx \, 
\frac{f_\pm(x,T,0)}{\sqrt{(x\pm\mu)^2 - \bar{m}^2(\nu)}} \, .
\hspace{5mm}
\label{pperpftanom}
\end{eqnarray}
In addition, when including the anomalous magnetic moment, the magnetization has a different form since
\begin{equation}
\bar{m}(\nu) \frac{\partial \bar{m}(\nu)}{\partial B} = 
-\bar{m}(\nu)\left[ s \kappa - \frac{|q| \nu}{\sqrt{m^2 + 2 \nu |q| B}} \right] ,
\end{equation}
which results in
\begin{eqnarray}
&&
M_\pm = \frac{P_{\parallel,\pm}}{B} + \frac{|q|B}{2\pi^2} \LLsumFT 
\bar{m}(\nu) \!\! \left[ s \kappa - \frac{|q| \nu}{\sqrt{m^2 + 2 \nu |q| B}} \right] 
\nonumber \\ &&
\hspace{3.0cm} \times
\int_{\bar{m}(\nu) \mp\mu}^{\infty} dx \, \frac{f_\pm(x,T,0)}{\sqrt{(x\pm\mu)^2 - \bar{m}^2(\nu)}}
\, .
\nonumber \\
\end{eqnarray}
Once again we see that $P_{\perp,\pm} = P_{\parallel,\pm} - M_\pm B$.

\section{Uncharged Particles}
\label{sec:uncharged}

In the case that the particle being considered is uncharged, then one does not obtain discrete 
Landau levels and, as a result,
\beq
\int_k \rightarrow \int  \frac{d^3k}{(2\pi)^3} \, ,
\eeq
in Eqs.~(\ref{ndens})-(\ref{pperp}).
Prior to proceeding with the 
calculations, we note that for uncharged particles one has
\begin{equation}
\bar{m}^2 = \left(\sqrt{m^2 + k_\perp^2} - s \kappa B \right)^2 \, .
\end{equation}

\subsection{Finite temperature}

We first consider the general case of uncharged particles at finite temperature including the
effect of the anomalous magnetic moment.  
The derivation necessary is performed in App.~\ref{app:tmunuderivation}.  Here we summarize
the results and list the contributions from particles and anti-particles.
The resulting expression for the number density is
\bqa
n_\pm &=& \frac{1}{2\pi^2} \sum_{s=\pm1} 
\int_{m - s \kappa B}^\infty dE \, E \, f_\pm(E,T,\mu)
\nonumber \\
&& \hspace{6mm} \times \left[ \hat{k} + s \kappa B 
\left(\arctan\left(\frac{s \kappa B - m}{\hat{k}}\right)
+\frac{\pi}{2}\right) \right] \! ,
\nonumber \\
\eqa
where $\hat{k} \equiv \sqrt{E^2 - (m - s\kappa B)^2}$.  The energy density is
\bqa
\epsilon_\pm &=& \frac{1}{2\pi^2} \sum_{s=\pm1} 
\int_{m - s \kappa B}^\infty dE \, E^2
\, f_\pm(E,T,\mu)
\nonumber \\
&& \hspace{6mm} \times \left[ \hat{k} + s \kappa B 
\left(\arctan\left(\frac{s \kappa B - m}{\hat{k}}\right)
+\frac{\pi}{2}\right) \right] .
\nonumber \\
\eqa
The longitudinal pressure is
\bqa
P_{\parallel,\pm} = &=& \frac{1}{24 \pi^2} \sum_{s=\pm1} \int_{m - s \kappa B}^\infty dE
\, f_\pm(E,T,\mu) 
\nonumber \\
&& \hspace{-2.1cm} \times
\biggl\{ 2 \hat{k} (s \kappa B - m)(2m + s \kappa B) 
\nonumber \\
&& \hspace{-1.8cm}
+ E^2 \left[ 4 \hat{k} + 6 s \kappa B \left(\arctan\left(\frac{s \kappa B - m}{\hat{k}}\right)
+\frac{\pi}{2}\right) \right] \biggr\} \, .
\nonumber \\
\eqa
The transverse pressure is
\bqa
P_{\perp,\pm} &=&  
\frac{1}{6\pi^2} \! \sum_{s=\pm1} \int_{m - s \kappa B}^\infty \! dE
\, f_\pm(E,T,\mu) 
( \hat{k}^3 - 3 s \kappa B m \hat{k} ) \, .
\nonumber \\
\eqa
Finally, we obtain the magnetization 
\bqa
M_\pm &=& \frac{\kappa}{4 \pi^2} \sum_{s=\pm1} s \int_{m - s \kappa B}^\infty dE
\, f_\pm(E,T,\mu) 
\nonumber \\
&& \hspace{-5mm} \times
\biggl[ \hat{k} (s \kappa B + m) + E^2
\left(\arctan\left(\frac{s \kappa B - m}{\hat{k}}\right) +\frac{\pi}{2}\right) 
\biggr] .
\nonumber \\
\eqa
We see that the magnetization vanishes when $\kappa \rightarrow 0$.  In addition,
with these expressions one finds $P_{\perp,\pm} = P_{\parallel,\pm} - M_\pm B$.

\subsection{Zero temperature}

In the zero temperature limit there is only a particle contribution since 
$\lim_{T\rightarrow0} f_-(E,T,\mu) = 0$ for $E\geq0$ 
and $\lim_{T\rightarrow0} f_+(E,T,\mu) =  \Theta(\mu - E)$.  Using the results listed in the 
previous subsection one finds for the number density \cite{Broderick:2000pe}
\begin{eqnarray}
&& n =  \frac{1}{4 \pi^2} \sum_{s=\pm1} \Biggl[
\frac{k_F}{3} \left( 2 k_F^2 - 3  s \kappa B \hat{m} \right) 
\nonumber \\
&& \hspace{2.0cm}
- s \kappa B \mu^2 \left( \arctan\left(\frac{\hat{m}}{k_F}\right) - \frac{\pi}{2} \right) \Biggr] \, ,
\end{eqnarray}
where $\hat{m} = m - s \kappa B$, $k_F = \sqrt{\mu^2 - \hat{m}^2}$.
Similarly the energy density can be obtained in this limit \cite{Broderick:2000pe}
\begin{eqnarray}
&&
\epsilon = 
\frac{1}{48\pi^2} \! \sum_{s=\pm1} \Biggl[
k_F \mu (6 \mu^2 - 3 \hat{m}^2 - 4 s \kappa B \hat{m})
\nonumber \\
&& \hspace{1cm}
- 8  s \kappa B \mu^3 \left( \arctan\left(\frac{\hat{m}}{k_F}\right) - \frac{\pi}{2} \right)
\nonumber \\
&& \hspace{1cm}
- \hat{m}^3 (3\hat{m} + 4 s \kappa B) \log\left(\frac{k_F + \mu}{\hat{m}}\right)
\Biggr] .
\end{eqnarray}
And the longitudinal pressure can also be easily obtained
\begin{eqnarray}
P_\parallel  &=& 
\frac{1}{48\pi^2} \! \sum_{s=\pm1} \Biggl[
k_F \mu (2 \mu^2 - 5 \hat{m}^2 - 8 s \kappa B \hat{m})
\nonumber \\
&& \hspace{1cm}
- 4  s \kappa B \mu^3 \left( \arctan\left(\frac{\hat{m}}{k_F}\right) - \frac{\pi}{2} \right)
\nonumber \\
&& \hspace{1cm}
+ \hat{m}^3 (3\hat{m} + 4 s \kappa B) \log\left(\frac{k_F + \mu}{\hat{m}}\right)
\Biggr] .
\label{zerotuppar}
\end{eqnarray}
Using the derived expressions for $n$, $\epsilon$, and $P_\parallel$ one can show that 
$\epsilon + P_\parallel = \mu n$ is satisfied explicitly.  The transverse pressure is
\bqa
P_\perp  &=& 
\frac{1}{48\pi^2} \! \! \sum_{s=\pm1} \! \biggl[
k_F \mu \left( 2\mu^2 - 5 \hat{m}^2 - 12 s \kappa B \hat{m} - 12 (s \kappa B)^2 \right)
\nonumber \\
&& \hspace{1.7cm}
+ 3 \hat{m}^2 (\hat{m}+2 s \kappa B)^2 \log\left(\frac{k_F + \mu}{\hat{m}}\right)
\biggr] .
\eqa
Finally, evaluating $\partial P_\parallel/\partial B$ one obtains the magnetization in this case 
\cite{Broderick:2000pe}\footnote{We note that there appear to be some typos in the expression contained
in Ref.~\cite{Broderick:2000pe}.}
\begin{eqnarray}
&&M = \frac{\kappa}{12\pi^2} \sum_{s=\pm1} s \biggl[
\mu k_F ( 3 s \kappa B + \hat{m} ) 
\nonumber \\ && \hspace{1.4cm}
- \mu^3 \left( \arctan\left(\frac{\hat{m}}{k_F}\right) - \frac{\pi}{2} \right)
\nonumber \\ && \hspace{1.4cm}
- \hat{m}^2 (3 s \kappa B  + 2 \hat{m}) \log\left(\frac{k_F + \mu}{\hat{m}}\right)
\biggr] .
\end{eqnarray}
From this result we can once again verify that $P_\perp = P_\parallel - M B$.

\section{Numerical results}
\label{sec:results}

In this section we present numerical evaluation of the transverse and longitudinal pressures derived in the
previous section.  For the numerics that follow we will assume (i) a gas of protons with a 
mass $m = m_p = 0.939$~GeV, electric charge $q = +e$, and an anomalous magnetic moment of 
$\kappa = \kappa_p  \mu_N = 1.79 \cdot e/(2 m_p) = 0.288633$ GeV$^{-1}$ in Heaviside-Lorentz 
natural units$\,$\footnote{In Gaussian 
natural units one has $\mu_N = 0.0454871$ GeV$^{-1}$ which is the 
Heaviside-Lorentz value divided by $\sqrt{4\pi}$.  Note that if one uses Gaussian natural units, the magnetic 
field in GeV$^2$ is scaled by a factor of $\sqrt{4\pi}$ compared to the corresponding Heaviside-Lorentz magnetic 
field.  As a result the product of $\mu_N B$ is independent of the convention chosen.} 
and (ii) a gas of neutrons with a mass $m=m_n = 0.939$ GeV, electric charge $q=0$, and an anomalous
magnetic momentum of $\kappa = \kappa_n  \mu_N = -1.91 \cdot e/(2 m_n) = -0.307983$ GeV$^{-1}$ 
\cite{Rabhi:2009ih}.
In all cases shown we consider a magnetic field magnitude of 
$5 \times 10^{18}$ Gauss.


\begin{figure}[t]
\includegraphics[width=0.45\textwidth]{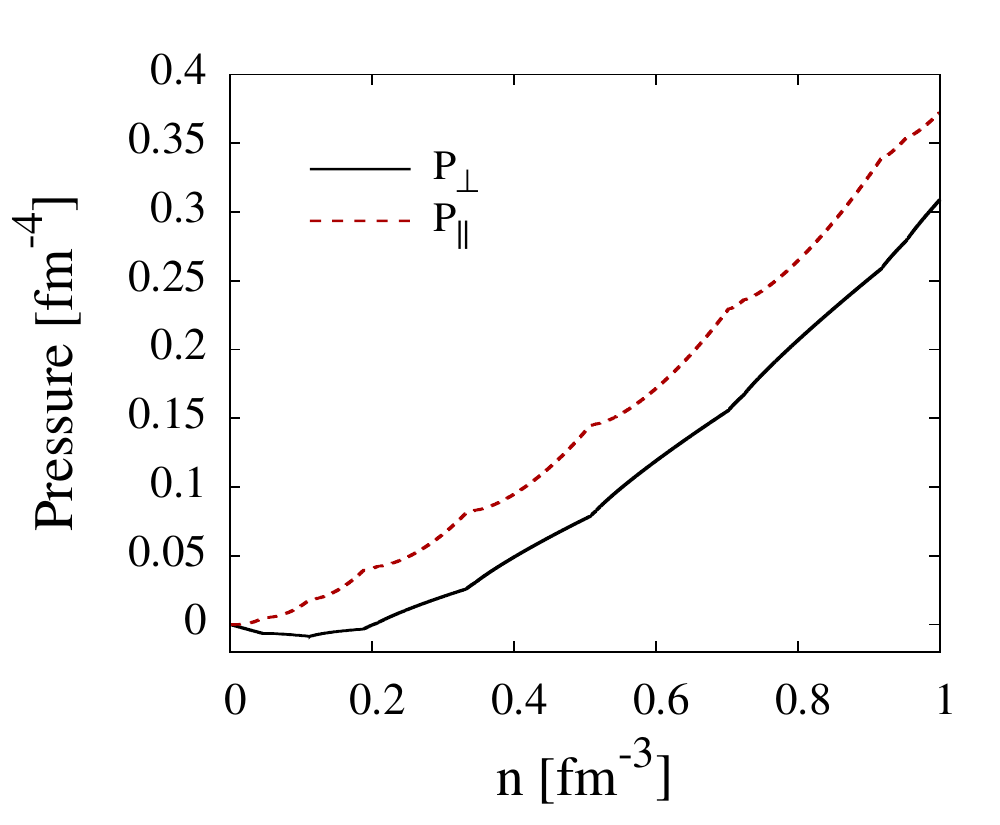}
\caption{(Color online) Transverse and longitudinal pressures of a zero temperature gas of protons 
as a function of the number density.  Results include the effect of the proton anomalous magnetic momentum.}
\label{fig:pkappa0}
\end{figure}

\begin{figure}[t]
\includegraphics[width=0.45\textwidth]{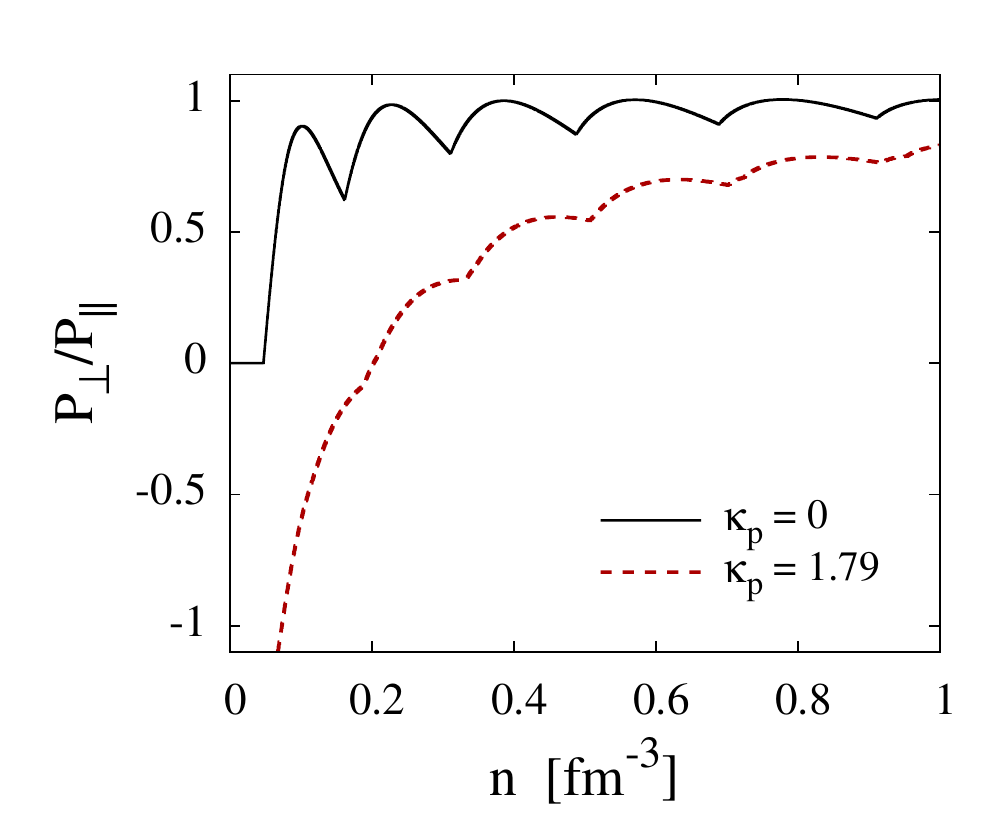}
\caption{(Color online) Ratio of transverse and longitudinal pressures of a zero temperature gas
of protons as a function of the number density.  Results are shown with and without the effect
of the proton anomalous magnetic moment.}
\label{fig:pkappacomp}
\end{figure}

\begin{figure}[t]
\includegraphics[width=0.45\textwidth]{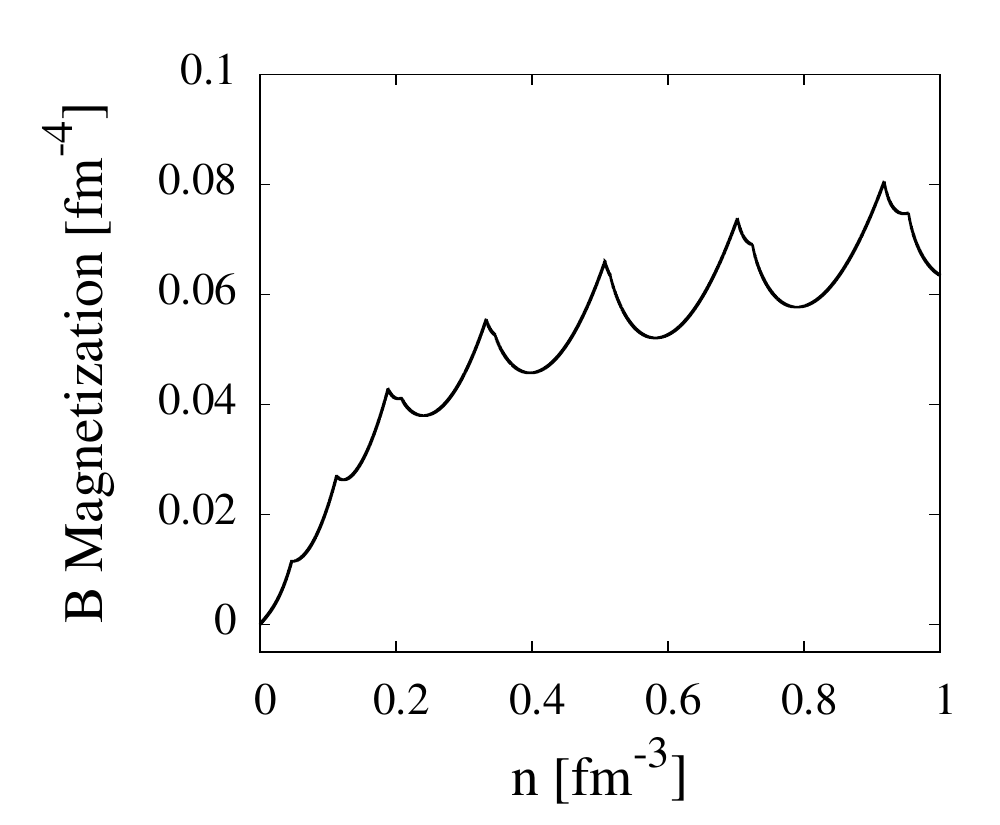}
\caption{ (Color online) Magnetization of a zero temperature gas of protons times the background
magnetic field.  Result includes the effect of the proton anomalous magnetic moment.}
\label{fig:mkappaf}
\end{figure}


In Fig.~\ref{fig:pkappa0} we plot the transverse and longitudinal pressures of a zero temperature gas of 
protons including the effect of the anomalous magnetic moment.   The cusps in the curves correspond to
threshold crossings for the maximum Landau level.  As can be seen from this figure, the transverse and 
longitudinal pressures are not equal.  
In addition, one can see from the figure that at low densities the
transverse pressure is negative at low densities when there is a non-vanishing anomalous magnetic 
moment, while the longitudinal pressure remains positive at all densities.

In Fig.~\ref{fig:pkappacomp} we show the ratio of the transverse to longitudinal pressures for a zero
temperature gas of protons with and without the effect of the anomalous magnetic moment. In both
cases we once again see cusps indicative of Landau level crossings and a vanishing transverse pressure 
at low densities.  From this figure we also see that including the anomalous magnetic moment enhances the pressure
anisotropy.

In Fig.~\ref{fig:mkappaf} we plot the background magnetic field times the magnetization of a zero 
temperature gas of protons obtained via Eq.~(\ref{cpmag}).
We note that there are two distinct sets of cusps visible in 
Fig.~\ref{fig:mkappaf}.  This is due to the fact that, when the effect of the anomalous
magnetic moment is included, there are two different Landau level thresholds for particles 
with spins aligned or anti-aligned with the background magnetic field.

\begin{figure}[t]
\includegraphics[width=0.45\textwidth]{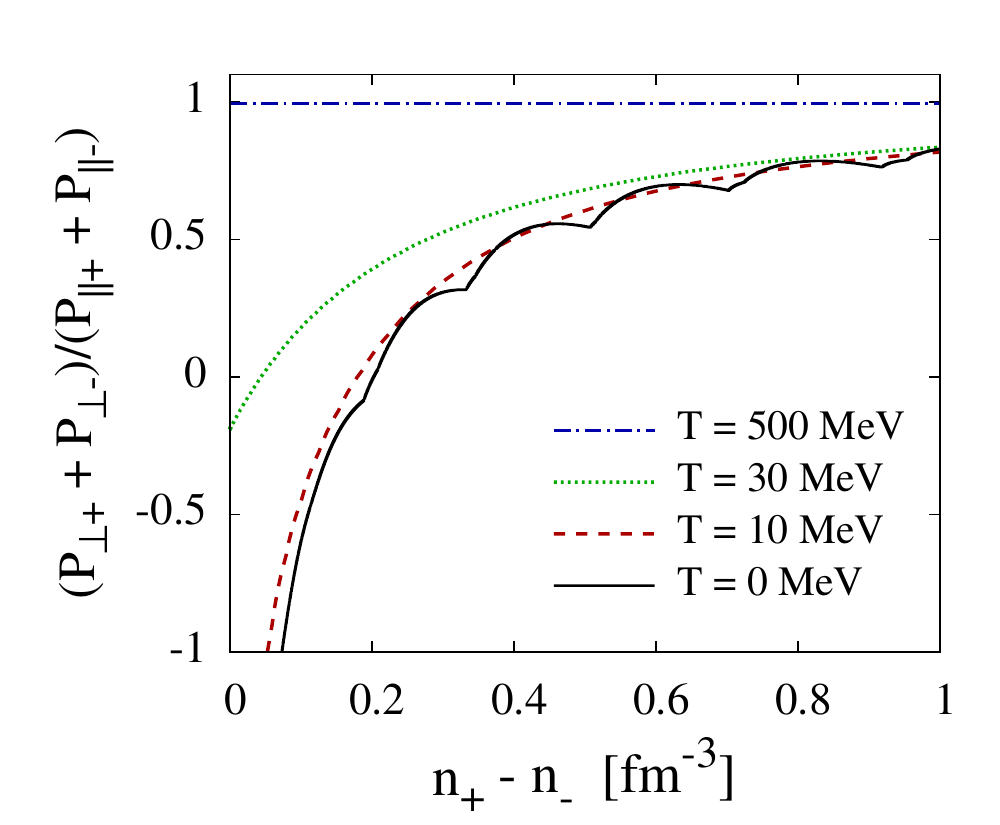}
\caption{(Color online) Ratio of transverse to longitudinal pressure of a gas of protons as a function of the 
net proton density for four different temperatures $T = \{0, 10, 30, 500\}$~MeV.  Results include the
effect of the proton anomalous magnetic moment.}
\label{fig:pkappaf}
\end{figure}

In Fig.~\ref{fig:pkappaf} we plot the ratio of the transverse pressure to the longitudinal pressure of a gas of 
protons as a function of the net proton density (particle minus anti-particle) for $T = \{0, 10, 30, 500\}$~MeV.  
As can be seen from this figure, as the temperature is increased, the cusps associated with
Landau level crossings are diminished and the level of the pressure anisotropy also decreases.  The 
highest temperature shown $T = 500$~MeV is on the order of those initially generated in relativistic
heavy ion collisions at CERN's Large Hadron Collider.  As we see, at these high temperatures the
pressure anisotropy for charged particles is quite small, $\lesssim 1\%$.  However, it should be noted
that as the system cools, the pressure anisotropy increases.

\begin{figure}[t]
\includegraphics[width=0.45\textwidth]{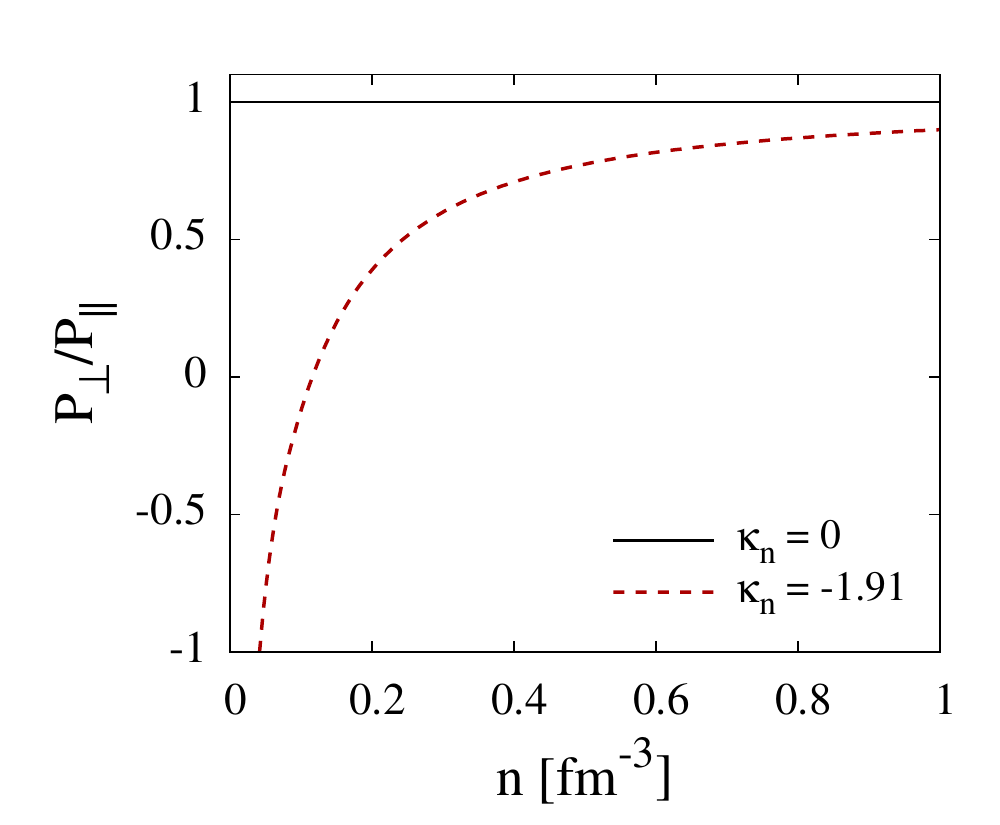}
\caption{(Color online) Ratio of transverse and longitudinal pressures of zero temperature gas of
neutrons as a function of  number density.  Results are shown with and without the effect of the neutron
anomalous magnetic moment.}
\label{fig:pkappaU}
\end{figure}

\begin{figure}[t]
\includegraphics[width=0.45\textwidth]{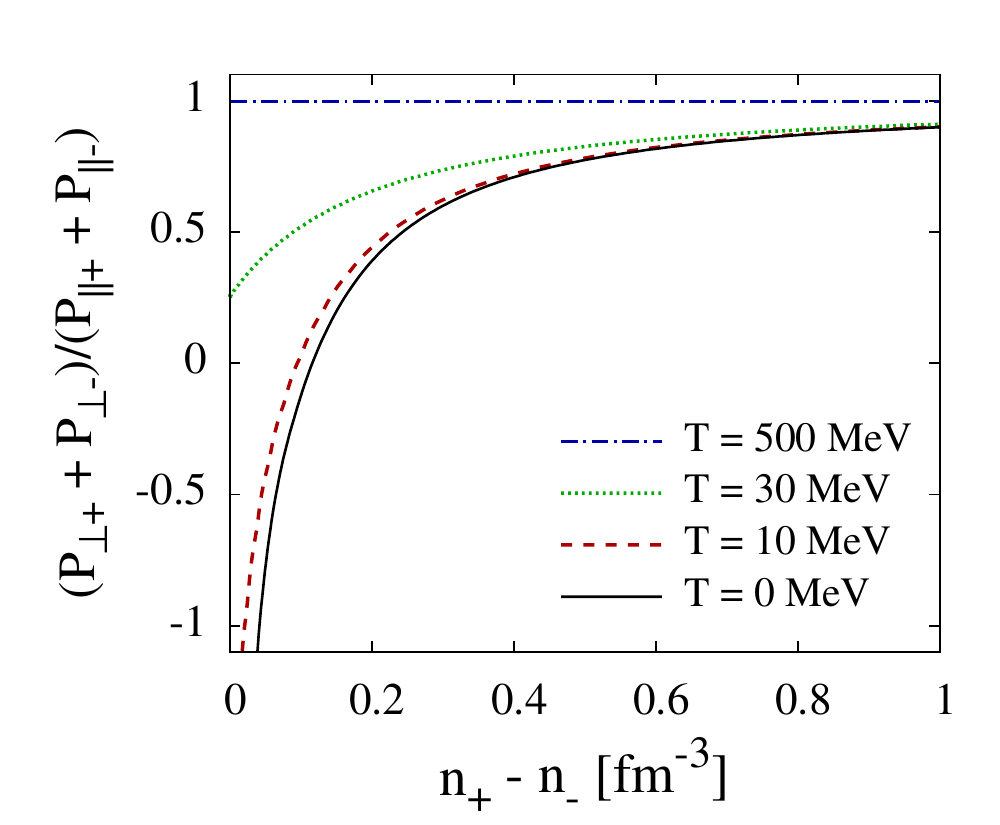}
\caption{(Color online) Ratio of transverse and longitudinal pressures of a gas of
neutrons as a function of the net neutron density for four different temperature $T = \{0, 10, 30, 500\}$~MeV.  
Results include the effect of the neutron anomalous magnetic moment.}
\label{fig:pkappaUT}
\end{figure}

We consider next the case of neutral particles, focusing on a specific example of a gas of neutrons.
In Fig.~\ref{fig:pkappaU} we plot the ratio of the transverse to longitudinal pressures of a gas
of neutrons as a function of the neutron density with and without the effect of the neutron anomalous
magnetic moment.  This figure shows that without the anomalous magnetic moment the pressures
are completely isotropic; however, when there is a non-vanishing anomalous magnetic moment the pressure
anisotropy can be quite sizable.  In Fig.~\ref{fig:pkappaUT} we show the ratio of the total particle plus 
anti-particle transverse to longitudinal pressures.  This figure shows that as the temperature of
the system increases, the amount of pressure anisotropy, again, decreases.


\section{Conclusions and outlook}
\label{sec:conclusions}

In this paper we have revisited the calculation of the matter contribution to the energy-momentum
tensor of a Fermi gas of spin one-half particles 
subject to an external magnetic field.  We considered both charged and uncharged
particles with and without the effect of the anomalous magnetic moment.  For zero temperature
systems we demonstrated through explicit calculation that the
resulting energy density, number density, and longitudinal pressure satisfy $\epsilon + P_\parallel  = 
\mu n$.   Using the standard definition of the grand potential $\Omega = \epsilon - \mu n$
allowed us to see that, in all cases investigated, the grand potential is related to the longitudinal 
pressure via $\Omega = - P_\parallel$ in agreement with previous findings in the literature.

We point out that some of the results contained herein are known in the literature.  
The results obtained for the transverse pressure of charged and uncharged particles with non-zero anomalous 
magnetic moment are new.  In addition, we have presented in two appendices an explicit derivation of the necessary
statistical averages of the energy-momentum tensor, taking into account the anomalous magnetic moment.  
Using the results obtained, we demonstrated that the standard relationship, $P_\perp = P_\parallel - M B$,  
between the transverse pressure, longitudinal pressure, and magnetization of the system holds in all cases 
considered.

The resulting formulae for the bulk properties can be applied to both zero temperature and finite temperature 
systems and hence could be useful in understanding the impact of high magnetic fields on the 
evolution of proto-neutron stars, proto-quark stars, and the matter generated in relativistic heavy ion 
collisions.  Applying the derived formulae to a system of protons we found that there can exist
a sizable pressure anisotropy in the matter contribution to the energy-momentum tensor which
could have a phenomenological impact.  Additionally we found that as the temperature of the
system increases, the pressure anisotropy decreases.  This is primarily due to 
the fact that increasing temperature allows higher Landau levels to be partially occupied and hence
reduces the discrete effects one sees at zero temperature.  For uncharged particles Landau 
quantization does not play a role and, instead, any pressure anisotropy exhibited comes from a non-vanishing
anomalous magnetic moment.  Once again as the temperature increases, the pressure anisotropy
is reduced.  This effect is due to the fact that as the temperature increases high momentum modes
become highly occupied which causes momentum terms in the energy to dominate over those 
associated with the anomalous magnetic moment.

We note that although we presented results applicable to the case of a single particle type, the resulting 
formulae can be easily applied to the case of a system composed of multiple particle types.  Since the 
contributing particles may have different pressure anisotropies depending on the sign
and the magnitude of the anomalous magnetic moment, one must 
take care to sum over all particle types subject to the necessary conservation laws prior to making 
quantitative statements about the phenomenological impact of magnetic-field induced pressure anisotropies 
on dense matter \cite{Dexheimer:2012mk}.  Finally, we emphasize that although the
numerical results shown in the results section assumed a particular magnetic field amplitude, the
analytic results derived herein are completely general and as such can be applied to assess the impact of
magnetic fields on the bulk properties of matter in a wide variety of situations.


\section*{Acknowledgments}

We thank Marcus Benghi Pinto for useful conversations.  We also thank an anonymous referee for useful feedback.
V.D. and D.P.M. acknowledge support from CNPq/Brazil.
M.S. thanks the Universidade Federal de Santa Catarina Florian\'opolis for hospitality during his stay in
Florian\'opolis.
M.S. was supported by NSF grant No.~PHY-1068765 and the Helmholtz International Center for FAIR 
LOEWE program.

\appendix

\section{Energy-momentum tensor}
\label{app:tmunubasics}

In this appendix we derive the energy-momentum tensor including the
effect of the anomalous magnetic moment.  For this purpose we will use the method of metric perturbations
which allows one to most efficiently compute a symmetric and gauge-invariant energy-momentum tensor.
The starting point is the following relation between the variation of the action and the energy momentum
tensor
\beq
\delta {\cal S} = \frac{1}{2} \int d^4x \, \sqrt{-g} \, {\cal T}^{\mu\nu}  \, \delta g_{\mu\nu} \, , 
\label{eq:svar}
\eeq
where $g \equiv {\rm det}(g_{\mu\nu})$.  We proceed in the standard way by 
writing the action in terms of the Lagrangian density, varying the metric, identifying
the energy-momentum tensor by comparison with (\ref{eq:svar}), and finally taking 
$g^{\mu\nu} \rightarrow  \eta^{\mu\nu}$ where $\eta^{\mu\nu} = {\rm diag}(1,-1,-1,-1)$ is the 
Minkowski-space metric.

We begin with the curved-space Lagrangian density for a spin one-half fermion with charge $q$ in the 
presence of an external magnetic field including the effect of the anomalous magnetic moment
\beq
{\cal L} = \bar\psi (i \slashed D - m  + \frac{1}{2} \kappa \sigma^{\mu\nu} F_{\mu\nu}) \psi 
- \frac{1}{4} F^{\mu\nu}F_{\mu\nu} \, ,
\label{eq:lagrangian}
\eeq
where $\kappa$ is the anomalous magnetic moment and, 
as usual, $\slashed a \equiv \gamma^\mu a_\mu$, $D_\mu \equiv \frac{1}{2}
( \overrightarrow{\partial}_{\!\!\mu} - \overleftarrow{\partial}_{\!\!\mu}) + \Gamma_\mu + i q A_\mu$
with $\Gamma_\mu$ being the spin connection which is zero in flat space, and 
$\sigma^{\mu\nu} = i [ \gamma^\mu,\gamma^\nu ]/2$.
This allows us to write the covariantized action as ${\cal S} =
\int d^4 x \, \sqrt{-g} \, {\cal L} = {\cal S}_m + {\cal S}_f $ with
\bqa
{\cal S}_m &=& \int d^4x \, \sqrt{-g} \, \bar\psi \biggl[
\frac{i}{2} \gamma^\alpha D^\beta (g_{\alpha\beta} + g_{\beta\alpha}) - m  
\nonumber \\
&& 
\hspace{4mm}
+ \frac{1}{8} \kappa \sigma^{\alpha\beta} F^{\gamma\delta} 
(g_{\alpha\gamma} + g_{\gamma\alpha}) (g_{\beta\delta} + g_{\delta\beta}) \biggr] \psi  \, , 
\hspace{4mm} \\
{\cal S}_f &=& - \frac{1}{4} \int d^4x \, \sqrt{-g} \, F^{\alpha\beta}F^{\gamma\delta} 
g_{\alpha\gamma} g_{\beta\delta} \, , 
\eqa
where we have split the action into matter and field contributions and used the the fact that the metric tensor is
symmetric to explicitly symmetrize the matter contribution.  First, we evaluate $\delta {\cal S}$ making use of the
identity $\delta \sqrt{-g} = -\frac{1}{2} \sqrt{-g} \, g^{\mu\nu} \delta g_{\mu\nu}$.
Note, importantly, that the gamma matrices
themselves depend on the metric and therefore one needs to take into account their
variation under metric variation.  The variation can be computed with \cite{wald1984} or without \cite{Weldon:2000fr}
the use of vierbeins.  Computing the variation, identifying ${\cal T}^{\mu\nu}$, and taking the limit $g^{\mu\nu} \rightarrow \eta^{\mu\nu}$ one finds the following expressions for the matter and field contributions to the
energy-momentum tensor in flat space
\bqa
{\cal T}_m^{\mu\nu} &=& \bar\psi \biggl[ \frac{i}{2} \left(\gamma^\mu D^\nu + \gamma^\nu D^\mu\right) 
\nonumber \\
&& \hspace{5mm} + \frac{1}{2} \kappa \left(\sigma^{\mu\alpha} {F^\nu}_{\!\alpha} + \sigma^{\nu\alpha} {F^\mu}_{\!\alpha} \right) \biggr] \psi - \eta^{\mu\nu} {\cal L}_m \, , 
\hspace{5mm}
\label{eq:tmnum}
\\
{\cal T}_f^{\mu\nu} &=& - F^{\mu\alpha} {F^\nu}_{\!\alpha} - \eta^{\mu\nu} {\cal L}_f \, ,
\label{eq:tmnuf}
\eqa
where ${\cal L}_m$ and ${\cal L}_f$ are the matter and field contributions to the Lagrangian density
corresponding to the first and second terms in Eq.~(\ref{eq:lagrangian}), respectively.  

\section{Matter contribution to $T^{\mu\nu}$}
\label{app:tmunuderivation}

In this appendix we derive expressions for the energy-momentum tensor in a uniform 
background magnetic field.  We focus on the matter contribution since the field contribution
(\ref{eq:tmnuf}) is standard.  In the rest of this appendix we can therefore ignore the pure gauge
field term in the Lagrangian.
In flat space the Lagrangian density for a spin one-half fermion with charge $q$ in the 
presence of an external magnetic field including the effect of the anomalous magnetic moment is
\beq
{\cal L} = \bar\psi (i \slashed D - m  + \frac{1}{2} \kappa \sigma^{\mu\nu} F_{\mu\nu}) \psi \, ,
\eeq
where $\kappa$ is the anomalous magnetic moment and, 
as usual, $\slashed a \equiv \gamma^\mu a_\mu$, $D_\mu \equiv \frac{1}{2}
( \overrightarrow{\partial}_{\!\!\mu} - \overleftarrow{\partial}_{\!\!\mu})  + i q A_\mu$, and 
$\sigma^{\mu\nu} = i [ \gamma^\mu,\gamma^\nu ]/2$.
The equations of motion for $\psi$ and $\bar\psi$ can be determined using
\bqa
&&\frac{\partial {\cal L}}{\partial \bar\psi} - 
\partial_\mu\!\left( \frac{\partial {\cal L}}{\partial (\partial_\mu \bar\psi)} \right) = 0 \, ,
\nonumber \\
&&\frac{\partial {\cal L}}{\partial \psi} - 
\partial_\mu\!\left( \frac{\partial {\cal L}}{\partial (\partial_\mu \psi)} \right) = 0 \, ,
\eqa
which result in 
\bqa
&&(i \slashed \partial - q \slashed A - m 
+ \frac{1}{2} \kappa \sigma^{\mu\nu} F_{\mu\nu} )\psi = 0  \, ,
\label{eq:eom1} \\
&&
i \partial_\mu \bar\psi \gamma^\mu + \bar\psi(q \slashed A + m 
- \frac{1}{2}\kappa \sigma^{\mu\nu} F_{\mu\nu}) = 0 \, .
\label{eq:eom2}
\eqa
We note for application to the calculation of $T^{\mu\nu}$ that if we multiply the first equation 
from the left by $\bar\psi$ we obtain ${\cal L}=0$.  This demonstrates that the matter Lagrangian density
vanishes when evaluated with solutions which obey the equations of motion.  This allows us to simplify Eq.~(\ref{eq:tmnum}) to
\bqa
{\cal T}^{\mu\nu} &=& \bar\psi \biggl[ \frac{i}{2} \left(\gamma^\mu D^\nu + \gamma^\nu D^\mu\right) 
\nonumber \\
&& \hspace{1cm} 
+ \frac{1}{2} \kappa \left(\sigma^{\mu\alpha} {F^\nu}_{\!\alpha} + \sigma^{\nu\alpha} {F^\mu}_{\!\alpha} \right) \biggr] \psi \, ,
\eqa

To evaluate  the necessary statistical average of ${\cal T}^{\mu\nu}$ we first need to solve the 
equations of motion in order to determine the energy eigenvalues and spinors.  The
spinors and energy eigenvalues are available in the literature 
\cite{ternov1966,Chand:1978bu,BagrovGitman1990,Melrose2013}; 
however, we review the derivation for sake of completeness and then use the resulting
spinors to evaluate 
the statistical averages of ${\cal T}^{\mu\nu}$.  We note that the spinor solutions have 
been expressed in various different forms in the literature.  We present a specific compact form
for the spinors, however, we have explicitly verified that using the forms of the spinors 
presented in Refs.~\cite{ternov1966,Chand:1978bu,BagrovGitman1990,Melrose2013} yields the
same final results.

As in the main body of the text, we choose the magnetic field to point along the $z$-direction.
Choosing the vector potential to be $A^\mu = (0,-By,0,0)$ we have $F^{\mu\nu} = B (\delta^{\mu x} 
\delta^{\nu y} - \delta^{\nu x} \delta^{\mu y})$ and as a result
\beq
\frac{1}{2} \kappa \sigma^{\mu\nu} F_{\mu\nu} = i \kappa B \gamma^x \gamma^y = 
\kappa B \left( \begin{array}{cc} \sigma_3 & 0 \\ 0 & \sigma_3 \end{array} \right) 
\equiv \kappa B {\cal S}_3 \, .
\eeq
Next we write Eq.~(\ref{eq:eom1}) in Hamiltonian form by searching for static solutions
of the form $\psi = e^{-i E t} \Psi({\bf x})$ which results in the Dirac-Pauli equation 
\cite{pauli1941}
\beq
( {\boldsymbol \alpha} \cdot {\boldsymbol \pi} 
+ \gamma^0 m - \kappa B \gamma^0 {\cal S}_3 ) \Psi = E \Psi \, ,
\label{eq:diracpaulieq}
\eeq
where ${\boldsymbol \alpha} \equiv \gamma^0 {\boldsymbol \gamma}$ and 
${\boldsymbol \pi} \equiv - i {\boldsymbol\nabla} - q {\bf A}$.  

Here we are interested in the diagonal components of ${\cal T}^{\mu\nu}$ which for a constant
magnetic field are given by
\bqa
{\cal T}^{00} &=& 
\bar\psi \left(i \gamma^0 D^0 \right) \psi \, ,
\\
{\cal T}^{xx} &=& 
\bar\psi \left(i \gamma^x D^x - \kappa B \sigma^{xy} \right) \psi \, ,
\\
{\cal T}^{yy} &=& 
\bar\psi \left(i \gamma^y D^y - \kappa B \sigma^{xy} \right) \psi \, ,
\\
{\cal T}^{zz} &=& 
\bar\psi \left(i \gamma^z D^z \right) \psi \, .
\eqa

\subsection{Charged particles}
\label{app:tmunuderivationcharged}

We now search for the solution of the Dirac-Pauli equation for charged particles.
Based on the structure of the equation, we begin by making an ansatz 
for the bi-spinor $\Psi$ of the form $\Psi({\bf x}) = e^{i k_x x} e^{i k_z z} u_n^{(s)}(y)$ 
with \cite{ternov1966}
\beq
u_n^{(s)}(y) = \left(
\begin{array}{l}
c_1 \phi_\nu(y) \\
c_2 \phi_{\nu-1}(y) \\
c_3 \phi_{\nu}(y) \\
c_4 \phi_{\nu-1}(y) \\
\end{array}
\right) ,
\eeq
where
\begin{equation}
\nu = n + \frac{1}{2} - \frac{s}{2} \frac{q}{|q|} \, ,
\end{equation}
with $n=0,1,2,\cdots$.\footnote{When $\nu=0$ there could be an issue with the Hermite
functions with index $\nu-1$ not being 
well defined; however, as we will show below, in this case one finds that the coefficients vanish identically.}
The constants $c_i$ above implicitly depend on the spin alignment $s=\pm1$. The functions 
$\phi_n$ are given by
\beq
\phi_n(\xi) = N_n e^{-\xi^2/2} H_n(\xi) \, ,
\eeq
where the variable $\xi$ is
\beq
\xi = \sqrt{|q|B} \left( y + \frac{k_x}{qB} \right) ,
\label{eq:xidef}
\eeq
$n \geq 0$ is an integer, 
$H_n$ is a Hermite polynomial, and $N_n = (q B)^{1/4}(\sqrt{\pi} 2^n n!)^{-1/2}$
is a normalization constant which ensures $\int_{-\infty}^\infty dy \, \phi_n^2(y)~=~1$.
Inserting this ansatz and simplifying the Dirac-Pauli equation, one obtains
\beq
\left(
\begin{array}{cccc}
m - \kappa B & 0 & k_z & k_\nu \\
0 & m + \kappa B & k_\nu & -k_z \\
k_z & k_\nu & -m+\kappa B & 0 \\
k_\nu & -k_z & 0 & -m - \kappa B \\
\end{array}
\right)
\chi = E 
\chi \, ,
\eeq
where $\chi = (c_1 \; c_2 \; c_3 \; c_4)^{\rm T}$ and $k_\nu = \sqrt{2 |q| B \nu}$.
Evaluating the determinant of the matrix on the left we obtain the energy
eigenvalues \cite{ternov1966}
\beq
E_s = \pm \sqrt{k_z^2 + (\lambda - s \kappa B)^2} \, , 
\eeq
where  
$\lambda \equiv \sqrt{m^2 + k_\nu^2}$.   The choice of an overall positive sign for
the energy eigenvalue above corresponds to particle states and the negative sign to 
anti-particle states.  Without loss of generality we can focus on the positive energy
states and, in the end, extend the result to include the necessary contribution from 
the negative energy states.

The resulting positive energy eigenvectors are
\beq
\chi^{(s)}
= \frac{1}{\sqrt{2\lambda\alpha_s\beta_s}} \left(
\begin{array}{c}
s \alpha_s \beta_s \\
-k_z k_\nu \\
s \beta_s k_z \\
\alpha_s k_\nu
\end{array}
\right) ,
\label{spinorsol}
\eeq
where $\alpha_s \equiv E_s - \kappa B  + s \lambda$ and $\beta_s \equiv  \lambda + s m$.
The overall normalization of the state is fixed by requiring that $\int_{-\infty}^\infty dy \, 
u^{(r)\dagger}_n({\bf x}) \, u^{(s)}_m({\bf x}) = 2 E_s \delta^{rs} \delta_{nm}$.
The general quantum state for positive energy states can now be constructed
\beq
\psi(x) = \sum_{s=\pm1} \sumint_k  \, b_s({\bf k}) u^{(s)}({\bf k}) e^{i {\tilde k}_\mu x^\mu} \, , 
\eeq
where $b_s({\bf k})$ is a particle creation operator which obeys
\beq
\{ b_r({\bf p}), b_s^\dagger({\bf k}) \} = (2 \pi) \delta_{rs} \delta_{nm} \delta(p_z - k_z) \, ,
\eeq
${\bf k} = (n,k_z)$ with $n=0,1,2,\cdots$, ${\tilde k} = (E_k,k_x,0,k_z)$, and
\beq
\sumint_k \equiv \frac{|q|B}{2\pi} \sum_n \int_{-\infty}^\infty \frac{d k_z}{2 \pi} \frac{1}{\sqrt{2 E_k}} \, .
\eeq
Note that the factor of $\sqrt{2 E_k}$ in the denominator above is fixed by the spinor normalization
used above.

\subsubsection{Energy Density}

To determine the energy density, we begin by evaluating the $00$ component of the energy-momentum 
density which is equivalent to the Hamiltonian density ${\cal T}^{00} = {\cal H} = i \psi^\dagger \partial_t \psi$.
Integrating over space gives the Hamiltonian
\bqa
H &=& i \int_x  \psi^\dagger \partial_t \psi \nonumber \\
&=& i \sum_{r,s} \int_x \sumint_p \sumint_k 
\left[ b_r^\dagger({\bf p}) u^{(r)\dagger}({\bf p}) e^{i {\tilde p}_\mu x^\mu}\right] 
\nonumber \\
&& \hspace{1cm}
\times \left[ b_s({\bf k}) u^{(s)}({\bf k}) (-i E_k) e^{-i {\tilde k}_\mu x^\mu} \right] 
,
\eqa
where $\int_x \equiv \int d^3 x$.
Using the orthonormality relations listed above one finds
\beq
H = \frac{|q|B}{2 \pi} \sum_{s=\pm1} \sum_n \int_{-\infty}^\infty \frac{d k_z}{2 \pi} 
E_k \, b_s^\dagger({\bf k}) b_s({\bf k})  \, . 
\eeq
We can now compute the thermal average of the energy using the density matrix $\rho$
\beq
\rho = e^{-\beta H + \alpha N} \, ,
\eeq
where is $\beta=1/T$ is the inverse temperature, 
$\alpha=\beta\mu$ with $\mu$ being the chemical potential, $H$ is the Hamiltonian operator, and
\beq
N = \frac{|q|B}{2 \pi} \sum_{s=\pm1} \sum_n \int_{-\infty}^\infty \frac{d k_z}{2 \pi} 
\, b^\dagger_s({\bf k})  b_s({\bf k}) \, ,
\eeq
is the number operator.  The statistical average of the 
Hamiltonian operator gives the energy density
\beq
\epsilon \equiv \langle H \rangle = \frac{\Tr[\rho H]}{\Tr[\rho]} \, .
\eeq
Using the Baker-Campbell-Hausdorff formula one obtains 
\beq
\langle b_s^\dagger({\bf k}) b_s({\bf k})\rangle = \langle b_s({\bf k}) b_s^\dagger({\bf k})
\rangle e^{- \beta (E - \mu)} \, ,
\eeq
which, upon application of the anti-commutation relations for the creation operators, gives 
the Fermi-Dirac distribution for particles
\beq
\langle b_s^\dagger({\bf k}) b_s({\bf k})\rangle = \frac{1}{e^{\beta(E_k-\mu)} +1} = f_+(E_k,T,\mu) \, .
\eeq
With this we obtain our final expression for the particle contribution to the energy density
\beq
\epsilon = \langle H  \rangle =  \frac{|q|B}{2 \pi} \sum_{s=\pm1} 
\sum_n \int_{-\infty}^\infty \frac{d k_z}{2 \pi} E_k f_+(E_k,T,\mu) \, .
\eeq
Note that if one includes the anti-particle states, one must normal order the Hamiltonian operator
prior to performing the statistical average.

\subsubsection{Number Density}

Based on the above discussion, the number density can easily be seen to be given by
\beq
n = \langle N  \rangle =  \frac{|q|B}{2 \pi} \sum_{s=\pm1} \sum_n \int_{-\infty}^\infty \frac{d k_z}{2 \pi}
 f_+(E_k,T,\mu) \, .
\eeq

\subsubsection{Longitudinal Pressure}

We now consider the longitudinal pressure which is given by $P_\parallel \equiv 
\langle \int_x {\cal T}^{zz} \rangle$ with
\bqa
{\cal T}^{zz} = \frac{i}{2}\left[\bar\psi\gamma^z \partial^z \psi 
- (\partial^z\bar\psi) \gamma^z \psi
\right] .
\eqa
Plugging in the explicit forms for the spinors we have
\bqa
\int_x {\cal T}^{zz}  &=& \frac{i}{2}
\sum_{r,s} \! \int_x \sumint_p \sumint_k \biggl\{
\left[ b_r^\dagger({\bf p}) u^{(r)\dagger}({\bf p}) e^{i {\tilde p}_\mu x^\mu}  \right] 
\nonumber \\
&& \hspace{1cm} \times \gamma^0 \gamma^z \left[ b_s({\bf k}) u^{(s)}({\bf k}) (ik^z) e^{-i {\tilde k}_\mu x^\mu} \right]
\nonumber \\
&& \hspace{0.6cm} - \left[ b_r^\dagger({\bf p}) u^{(r)\dagger}({\bf p}) (-ip^z) e^{i {\tilde p}_\mu x^\mu} \right] 
\nonumber \\
&& \hspace{1cm} \times \gamma^0 \gamma^z \left[ b_s({\bf k}) u^{(s)}({\bf k}) e^{-i {\tilde k}_\mu x^\mu} \right] \biggr\} .
\eqa
Evaluating the $x$ and $p$ (sum-)integrals, making use of the orthonormality relations 
and then taking the statistical average gives
\bqa
P_\parallel &=& -\frac{1}{2} 
\frac{|q|B}{2 \pi} \sum_{s=\pm1} \sum_n \int_{-\infty}^\infty \frac{d k_z}{2 \pi}
\frac{k^z}{E_k}
\langle b_s^\dagger({\bf k}) b_s({\bf k})  \rangle
\nonumber \\
&& \hspace{1.3cm} \times 
\int_{-\infty}^\infty dy \, [ u^{(s)\dagger}({\bf k}) \gamma^0 \gamma^z u^{(s)}({\bf k}) ]
\, ,
\hspace{4mm}
\eqa
where we have used the fact that $\langle b_r^\dagger({\bf k}) b_s({\bf k})  \rangle$ vanishes unless
$r=s$.  Next we need to evaluate the spinor contraction
\beq
\int_{-\infty}^\infty dy \, u^{(s)\dagger}({\bf k}) \gamma^0 \gamma^z u^{(s)}({\bf k}) = - 2 k^z \, ,
\eeq
which follows from the explicit form of the spinors obtained previously.  Using this and rewriting the
statistical average of the number operator as a Fermi-Dirac distribution, one obtains
\beq
P_\parallel =  
\frac{|q|B}{2 \pi} \sum_{s=\pm1} \sum_n \int_{-\infty}^\infty \frac{d k_z}{2 \pi}
\frac{k^z k^z}{E_k} f_+(E_k,T,\mu) \, .
\eeq

\subsubsection{Transverse Pressure}

We finally turn our attention to the transverse pressure.  By rotational symmetry, 
$P_\perp \equiv \langle \int_x {\cal T}^{yy} \rangle = \langle \int_x {\cal T}^{xx} \rangle $.
Choosing the former, which is somewhat easier to evaluate, we should integrate and statistically
average
\beq
{\cal T}^{yy} = \frac{i}{2}\left[\bar\psi\gamma^y \partial^y \psi 
- (\partial^y\bar\psi) \gamma^y \psi \right] - \kappa B \bar\psi \sigma^{xy} \psi .
\eeq
Plugging in the explicit forms for the spinors we have
\bqa
\int_x {\cal T}^{yy}  &=& \frac{i}{2}
\sum_{r,s} \! \int_x \sumint_p \sumint_k \biggl\{
\left[ b_r^\dagger({\bf p}) u^{(r)\dagger}({\bf p}) e^{i {\tilde p}_\mu x^\mu}  \right] 
\nonumber \\
&& \hspace{1cm} \times \gamma^0 \gamma^y \left[ b_s({\bf k}) \partial^y u^{(s)}({\bf k}) e^{-i {\tilde k}_\mu x^\mu} \right]
\nonumber \\
&& \hspace{0.6cm} - \left[ b_r^\dagger({\bf p}) \partial^y u^{(r)\dagger}({\bf p}) e^{i {\tilde p}_\mu x^\mu} \right] 
\nonumber \\
&& \hspace{1cm} \times \gamma^0 \gamma^y \left[ b_s({\bf k}) u^{(s)}({\bf k}) e^{-i {\tilde k}_\mu x^\mu} \right] \biggr\}
\hspace{7mm} \nonumber
\\
&& - \kappa B \sum_{r,s} \! \int_x \sumint_p \sumint_k
\left[ b_r^\dagger({\bf p}) u^{(r)\dagger}({\bf p}) e^{i {\tilde p}_\mu x^\mu}  \right]  
\nonumber \\
&& \hspace{1cm} \times
\gamma^0 
\sigma^{xy} 
\left[ b_s({\bf k}) u^{(s)}({\bf k}) e^{-i {\tilde k}_\mu x^\mu} \right] .
\eqa
Evaluating the $x$ and $p$ (sum-)integrals making use of the orthonormality relations 
and then taking the statistical average gives
\bqa
P_\perp &=& 
\frac{|q|B}{2 \pi} \sum_{s=\pm1} \sum_n \int_{-\infty}^\infty \frac{d k_z}{2 \pi}
\frac{1}{E_k} \langle b_s^\dagger({\bf k}) b_s({\bf k})  \rangle
\nonumber \\
&& \hspace{5mm} \times \biggl\{
\frac{i}{2} \int_{-\infty}^\infty dy \biggl[ u^{(s)\dagger}({\bf k}) \gamma^0 \gamma^y \partial^y u^{(s)}({\bf k}) 
\nonumber \\
&& \hspace{3cm}
- \partial^y u^{(s)\dagger}({\bf k}) \gamma^0 \gamma^y u^{(s)}({\bf k}) \biggr] 
\nonumber \\
&& \hspace{1cm}
- \kappa B \int_{-\infty}^\infty dy \, u^{(s)\dagger}({\bf k})  \gamma^0 
\sigma^{xy} u^{(s)}({\bf k}) 
\biggr\} .
\hspace{8mm}
\eqa
Integrating by parts one finds that the second term contributes the same as the first.
Using the explicit representation of the spinors obtained above one finds
\bqa
&&\int_{-\infty}^\infty dy \, u^{(s)\dagger}({\bf k}) \gamma^0 \gamma^y \partial^y u^{(s)}({\bf k})
\nonumber \\
&& \hspace{1cm}
= -i (c_2 c_3 - c_1 c_4) 
\nonumber \\
&& \hspace{2cm}
\times \int_{-\infty}^\infty d\xi \left( \phi_{\nu-1} \partial_\xi \phi_{\nu} - \phi_{\nu} \partial_\xi \phi_{\nu-1} \right) 
\nonumber \\
&& \hspace{1cm}
= -i \sqrt{2 |q| B \nu} (c_2 c_3 - c_1 c_4) 
\nonumber \\
&& \hspace{1cm}
= -i 2 |q| B \nu \left(1 - \frac{s \kappa B}{\lambda}\right) ,
\eqa
and
\beq
\int_{-\infty}^\infty dy \, u^{(s)\dagger}({\bf k})  \gamma^0 
\sigma^{xy} u^{(s)}({\bf k}) = 2 s  (\lambda -  s \kappa B) \, .
\eeq
With this we can write down our final expression for the transverse pressure for charged particles
\bqa
P_\perp &=&  
 \frac{|q| B^2}{2 \pi^2} \sum_{s=\pm1} \sum_n \int_{-\infty}^\infty d k_z
\frac{1}{E_k} f_+(E_k,T,\mu)
\nonumber \\
&& \hspace{1.5cm} 
\times \Biggl[ 
\frac{|q| \nu \bar{m}(\nu)}{\sqrt{m^2 + 2 \nu |q| B}} -   s \kappa \bar{m}(\nu) \Biggr] .
\hspace{4mm}
\eqa
where $\bar{m}(\nu) \equiv \sqrt{m^2 + 2 \nu |q| B} - s \kappa B$.

\subsection{Uncharged Particles}
\label{app:tmunuderivationuncharged}

We now consider the case of uncharged particles.  This case is different since the transverse
momenta of the particles are not quantized.  Starting from the Dirac-Pauli equation
(\ref{eq:diracpaulieq}) we make an ansatz for the bi-spinor $\Psi$ of the form $\Psi({\bf x}) = 
e^{i {\bf k} \cdot {\bf x}} u$ with $u = (c_1 \; c_2 \; c_3 \; c_4)^{\rm T}$.
This results in the following matrix equation for $u$
\beq
\left(
\begin{array}{cccc}
m - \kappa B & 0 & k_z & k_- \\
0 & m + \kappa B & k_+ & -k_z \\
k_z & k_- & -m+\kappa B & 0 \\
k_+ & -k_z & 0 & -m - \kappa B \\
\end{array}
\right)
u = E 
u \, ,
\eeq
where $k_\pm \equiv k_x \pm i k_y$.
Evaluating the determinant of the matrix on the left we obtain the energy
eigenvalues
\beq
E_s = \pm \sqrt{k_z^2 + (\lambda - s \kappa B)^2} \, , 
\eeq
where now we have
$\lambda \equiv \sqrt{m^2 + k_\perp^2}$ with $k_\perp^2 = k_x^2 + k_y^2$. 
Once again the choice of an overall positive sign corresponds to particle states and negative
sign to anti-particle states.  We focus on particle states since the result is straightforward
to extend to anti-particles.

The resulting positive energy solutions are
\beq
u^{(s)}
= \frac{1}{\sqrt{2\lambda\alpha_s\beta_s}} \left(
\begin{array}{c}
s \alpha_s \beta_s \\
-k_z k_+ \\
s \beta_s k_z \\
\alpha_s k_+
\end{array}
\right) ,
\eeq
where as before $\alpha_s \equiv E_s - \kappa B  + s \lambda$ and $\beta_s \equiv  \lambda + s m$.
The overall normalization of the state is fixed in this case 
by requiring that $u^{(r)\dagger} \, u^{(s)} = 2 E_s \delta^{rs}$.
The general quantum state for positive energy states can now be constructed
\beq
\psi(x) = \sum_{s=\pm1} \int_k \frac{1}{\sqrt{2 E_k}}  \, b_s({\bf k}) u^{(s)}({\bf k}) e^{i k_\mu x^\mu} \, , 
\eeq
where $b_s({\bf k})$ is a particle creation operator and $\int_k = (2\pi)^{-3} \int d^3k$.  
Once again the factor of $\sqrt{2 E_k}$ in the denominator above is fixed by the spinor 
normalization used above.

Following the same general procedures used in the charged particle derivation one obtains 
the following result for the energy density
\beq
\epsilon = \langle H  \rangle =  \sum_{s=\pm1} 
\int_k E_k f_+(E_k,T,\mu) \, .
\eeq
The result for the number density is
\beq
n = \langle N  \rangle =  \sum_{s=\pm1} 
\int_k f_+(E_k,T,\mu) \, .
\eeq
The result for the parallel pressure is
\beq
P_\parallel = \langle T^{zz} \rangle = \sum_{s=\pm1} 
\int_k \frac{k_z^2}{E_k} f_+(E_k,T,\mu) \, .
\eeq
And, finally, the result for the transverse pressure $P_\perp = \langle T^{xx} \rangle = \langle T^{yy}\rangle$ is
\beq
P_\perp =  
\sum_{s=\pm1} 
\int_k \frac{1}{E_k} 
\Biggl[ 
\frac{1}{2} \frac{k_\perp^2 \bar{m}}{\sqrt{m^2 + k_\perp^2}}
-   s \kappa B \bar{m}
 \Biggr]
 f_+(E_k,T,\mu) \, .
\eeq
where $\bar{m} = \sqrt{m^2 + k_\perp^2} - s \kappa B$.

In all of the expressions above, we can perform two of the three integrations by making the following change of 
variables
\bqa
k_x &=& \sqrt{\lambda^2 - m^2}\cos\phi \, , \nonumber \\
k_y &=& \sqrt{\lambda^2 - m^2}\sin\phi \, , \nonumber \\
k_z &=&  \sqrt{E^2 - (\lambda - s \kappa B)^2} \, .
\eqa
Evaluating the Jacobian one finds
\beq
d^3k = \frac{E \lambda}{\sqrt{E^2 - (\lambda - s \kappa B)^2}} \, dE \, d\lambda \, d\phi \, .
\eeq
With this change of variables we obtain the number density
\bqa
n &=& \frac{1}{2\pi^2} \sum_{s=\pm1} 
\int_{m - s \kappa B}^\infty dE \, E
\, f_+(E,T,\mu) 
\nonumber \\
&& \hspace{1.5cm} \times \int_m^{E + s \kappa B} d\lambda 
\frac{\lambda}{\sqrt{E^2 - (\lambda - s \kappa B)^2}}
\nonumber \\
&=& \frac{1}{2\pi^2} \sum_{s=\pm1} 
\int_{m - s \kappa B}^\infty dE \, E \, f_+(E,T,\mu)
\nonumber \\
&& \hspace{6mm} \times \left[ \hat{k} + s \kappa B 
\left(\arctan\left(\frac{s \kappa B - m}{\hat{k}}\right)
+\frac{\pi}{2}\right) \right] \! ,
\nonumber \\
\eqa
where $\hat{k} \equiv \sqrt{E^2 - (m - s\kappa B)^2}$.  The energy density is given by
\bqa
\epsilon &=& \frac{1}{2\pi^2} \sum_{s=\pm1} 
\int_{m - s \kappa B}^\infty dE \, E^2
\, f_+(E,T,\mu) 
\nonumber \\
&& \hspace{1.5cm} \times \int_m^{E + s \kappa B} d\lambda 
\frac{\lambda}{\sqrt{E^2 - (\lambda - s \kappa B)^2}}
\nonumber \\
&=& \frac{1}{2\pi^2} \sum_{s=\pm1} 
\int_{m - s \kappa B}^\infty dE \, E^2
\, f_+(E,T,\mu)
\nonumber \\
&& \hspace{6mm} \times \left[ \hat{k} + s \kappa B 
\left(\arctan\left(\frac{s \kappa B - m}{\hat{k}}\right)
+\frac{\pi}{2}\right) \right] .
\nonumber \\
\eqa
For the parallel pressure we obtain
\bqa
P_\parallel &=& \frac{1}{2\pi^2} \sum_{s=\pm1} \int_{m - s \kappa B}^\infty dE
\, f_+(E,T,\mu) 
\nonumber \\
&& \hspace{5mm} \times \int_m^{E + s \kappa B} d\lambda \, 
\lambda \sqrt{E^2 - (\lambda^2 - s \kappa B)^2} \, ,
\nonumber \\
&=& \frac{1}{24 \pi^2} \sum_{s=\pm1} \int_{m - s \kappa B}^\infty dE
\, f_+(E,T,\mu) 
\nonumber \\
&& \hspace{0.5cm} \times
\biggl\{ 2 \hat{k} (s \kappa B - m)(2m + s \kappa B) 
\nonumber \\
&& \hspace{-0.2cm}
+ E^2 \left[ 4 \hat{k} + 6 s \kappa B \left(\arctan\left(\frac{s \kappa B - m}{\hat{k}}\right)
+\frac{\pi}{2}\right) \right] \biggr\} \, ,
\nonumber \\
\eqa
and for the perpendicular pressure we obtain
\bqa
P_\perp &=&  
\frac{1}{2\pi^2} \sum_{s=\pm1} \int_{m - s \kappa B}^\infty dE
\, f_+(E,T,\mu) \int_m^{E + s \kappa B} d\lambda \, 
\nonumber \\
&& 
\times 
\frac{\lambda - s \kappa B}{\sqrt{E^2 - (\lambda^2 - s \kappa B)^2}}
\Biggl[ 
\frac{1}{2} (\lambda^2 - m^2) 
-   s \kappa B \lambda
\Biggr] ,
\nonumber \\
&=&  
\frac{1}{6\pi^2} \sum_{s=\pm1} \int_{m - s \kappa B}^\infty dE
\, f_+(E,T,\mu) 
( \hat{k}^3 - 3 s \kappa B m \hat{k} ) \, .
\nonumber \\
\eqa

\vspace{2cm}


\bibliography{anisomatter}

\end{document}